\documentstyle[prb,aps,twocolumn,epsf]{revtex}
\begin{document}
\draft
\preprint{cond-mat}
\twocolumn[\hsize\textwidth\columnwidth\hsize\csname @twocolumnfalse\endcsname

\title{Structural and superconducting transition in selenium under high pressures}
\author{M. Otani\cite{byline}}
\address{Division of Material Physics, Department of Physical Science, Graduate School of Engineering Science, \\Osaka University, 1-3 Machikaneyama-cho, Toyonaka 560--8531, Japan}
\author{N. Suzuki}
\address{Division of Material Physics, Department of Physical Science, Graduate School of Engineering Science, \\Osaka University, 1-3 Machikaneyama-cho, Toyonaka 560--8531, Japan\\
and ``Research Area'' CREST, Japan Science and Technology Corporation}
\date{\today}
\maketitle
\begin{abstract}
	First-principles calculations are performed for electronic structures 
of two high pressure phases of solid selenium, $\beta$-Po and bcc. 
	Our calculation reproduces well the pressure-induced phase 
transition from $\beta$-Po to bcc observed in selenium.
	The calculated transition pressure is 30 GPa lower than the 
observed one, but the calculated pressure dependence of the lattice parameters 
agrees fairly well with the observations in a wide range of pressure. 
	We estimate the superconducting transition temperature 
$T_{\rm c}$ of both the $\beta$-Po and the bcc phases by calculating 
the phonon dispersion and the electron-phonon interaction on the basis of 
density-functional perturbation theory. 
	The calculated $T_{\rm c}$ shows a characteristic pressure dependence, 
{\it i.e.} it is rather pressure independent in the $\beta$-Po phase, 
shows a discontinuous jump at the transition from $\beta$-Po to bcc, 
and then decreases rapidly with increasing pressure in the bcc phase. 
\end{abstract}

\vskip2pc]

\narrowtext
\section{INTRODUCTION}
	Recently, with the development of high-pressure experimental techniques, 
research on pressure-induced structural phase transitions of the 
group VIb elements O, S, Se and Te has progressed greatly.  
	At high-pressures both metallic tellurium (Te) and 
selenium (Se) transform from $\beta$-Po to bcc phase at 27~GPa~\cite{rf:1} 
and 150~GPa,\cite{rf:2}, respectively.
	The $\beta$-Po type structure is rhombohedral and can be described 
as a simple cubic lattice deformed along the [111] direction keeping 
the edge length unchanged. 
	By changing the ratio $c/a$ of the rhombohedral lattice we obtain the 
bcc structure when $c/a=\sqrt{6}/4$. 
	In Te superconducting transitions are observed in both $\beta$-Po 
and bcc structures at low temperature and a jump in $T_{\rm c}$ 
from 2.5~K to 7.4~K is observed at 32--35~GPa.\cite{rf:2} 
	Theoretically, Mauri, {\it et}. {\it al}. suggests that the jump in 
$T_{\rm c}$ is related to the phonon softening in the bcc phase, 
namely with decreasing pressure the phonon anomaly enhances the 
electron-phonon coupling.\cite{44} 
        In Se, on the other hand, there is neither experimental observation 
nor {\it ab} {\it initio} calculation for the pressure dependence of 
phonon frequencies, the electron-phonon interaction or $T_{\rm c}$.

	The purpose of the present paper is to estimate these quantities by 
using {\it ab} {\it initio} calculations. 
        First we calculate the total energies of the $\beta$-Po and bcc phases 
of Se by using the full-potential linearized muffin-tin orbital (FPLMTO) method~\cite{05} 
in order to discuss the pressure-induced phase transition from 
$\beta$-Po to bcc. 
	Then we calculate phonon frequencies and electron-phonon coupling 
constants using the linear-response FPLMTO(LR-FPLMTO) 
method.\cite{06,13} 
        The validity of this method is demonstrated in references  5--8. 
	Finally we calculate the superconducting transition temperature 
in both the $\beta$-Po and bcc phases as a function of pressure. 

\section{STRUCTRAL TRANSITION FROM $\beta$-Po to bcc IN SELENIUM}
\subsection{Calculational procedure}

	The calculations of electronic states for the $\beta$-Po and bcc structures of Se  have been done according to the following procedure with use of the FPLMTO program.
	For exchange-correlation functional we have adopted the formula proposed by Gunnarsson and Lundqvist~\cite{10} and the GGA correction proposed by Perdew {\it et. al.}\cite{11} has been taken into account.
	Inside the muffin-tin (MT) spheres the scalar-relativistic calculations are performed for valence electrons, and the core states are recalculated at each self-consistent iteration with relativistic effects. 
	The MT radius has been taken to be 1.07~\AA. 
	The $\bbox{k}$-space integration has been performed by the improved tetrahedron method~\cite{12} with use of (12, 12, 12) grid of the sampling $\bbox{k}$-points [189 points in the IBZ ]. 
	We have used 3$\kappa$--$spd$--LMTO basis set (27 orbitals): $\kappa^2=-$0.1 ,$-$1.0 and $-$2.0~Ryd. 
	In the interstitial region the basis functions are expanded in plane waves up to the cutoff corresponding approximately to 200, 350 and 650 plane waves per $s$, $p$, and $d$ orbitals, respectively. 
	The charge densities and the potentials are expanded inside the MT spheres by spherical harmonics up to $\ell_{\rm max}$=6 and in the interstitial region by plane waves with the cutoff corresponding to the (16, 16, 16) fast-Fourier-transform(FFT) grid in the unit cell of direct space.
	The final convergence is within 10$^{-6}$~Ryd. 

\subsection{Results}
\begin{figure}[h]
\begin{center}
\leavevmode
\epsfxsize=80mm
\epsfbox{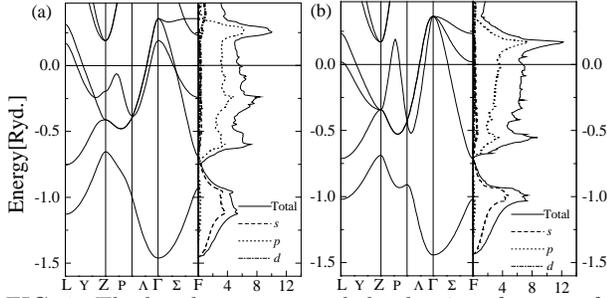}
\caption{The band structure and the density of states of Se at atomic volume $V$=14.82~\AA: (a) $\beta$-Po type structure and (b) bcc structure. The horizontal line denotes the Fermi level.}
\label{fig:01}
\end{center}
\end{figure}

	As the first step to investigate the phase transition from $\beta$-Po to bcc we have calculated the electronic band structure and the total energy of 
$\beta$-Po Se at atomic volume $V_{{\rm A}}$=14.82\AA. 
	At this volume the $\beta$-Po structure is known to be stable by experimental measurements. 
        For comparison we have calculated the electronic band structure and the total energy of the hypothetical bcc structure with the same atomic volume.
	The calculated energy band structures and the DOS for the $\beta$-Po and the bcc structures are shown in Fig.~\ref{fig:01}.
        The lowest band is mainly derived from the 4$s$ component and the next three bands from the 4$p$ component.
	The band structures of both the structures are similar to each other on the whole. 
	However, remarkable differences can be seen in the band structure and the DOS near the Fermi level. 
        Firstly, the band structure of the bcc structure along the P and $\Lambda$ lines has much larger dispersion than that of the $\beta$-Po type structure. 
        Secondly, the DOS at the Fermi level of the bcc structure is larger than that of the $\beta$-Po structure because in bcc Se the L point is a saddle point of the third energy band from the bottom which crosses the Fermi level near the L point. 
	By deforming from bcc to $\beta$-Po the L point energy of the third band goes up away from the Fermi level and the states of the fourth band in the middle of the P line come down under the Fermi level. 
       As a result the DOS of $\beta$-Po Se has a relatively large peak at 0.225 Ryd. below the Fermi level.
       This is the reason why the $\beta$-Po type structure is relatively more stable compared with the bcc structure. 

\begin{figure}[h]
\begin{center}
\leavevmode
\epsfxsize=80mm
\epsfbox{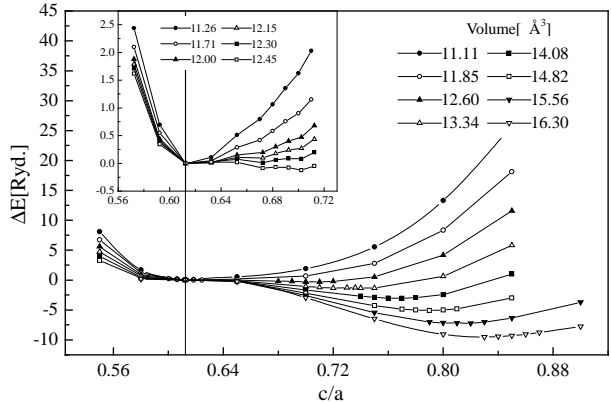}
\caption{The total energy of $\beta$-Po Se as a function of c/a for the fixed atomic volume. The inset indicate the results for volumes near the phase transition. All the energies are referenced to the bcc structure, {\it i.e.} $\Delta E$=$E$-$E_{{\rm bcc}}$.}
\label{fig:02}
\end{center}
\end{figure}

	To investigate the pressure-induced structural transition we have 
to calculate the total energy of the $\beta$-Po and the bcc structures 
as a function of volume. 
	For the $\beta$-Po structure we have optimized $c/a$ at each volume, 
namely, we have calculated the total energy of the $\beta$-Po type Se 
as a function of $c/a$ with the atomic volume $V_{\rm A}$ being kept constant. 
	Figure~\ref{fig:02} shows the total energy {\it vs} $c/a$ 
for several fixed atomic volume and the inset shows the results 
for a volume range near the phase transition. 
        All the energies are referenced to that of the bcc structure, 
{\it i.e.} the energy at $c/a$=$\sqrt{6}/4$. 
        For a large volume such as 16.30~\AA$^3$ the energy takes the minimum 
at $c/a$ $\sim$ 0.85 and the $c/a$ of bcc is an inflection point. 
        With decreasing volume the value of the $c/a$ starts to decrease gradually toward the bcc structure. 
        Finally, the $\beta$-Po structure is no longer a quasistable structure, that is, the total energy has a single minimum at $c/a$=$\sqrt{6}/4$ corresponding to the bcc structure.

\begin{figure}[h]
\begin{center}
\leavevmode
\epsfxsize=80mm
\epsfbox{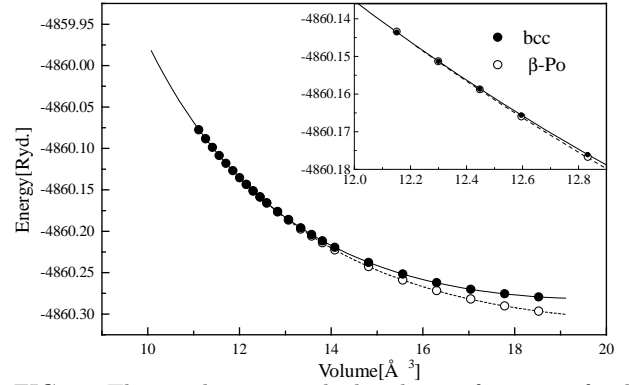}
\caption{The total energy calculated as a function of volume for the $\beta$-Po and 
the bcc structures of Se. The lines are obtained by using Murnanghan's equation of 
state.}
\label{fig:03}
\end{center}
\end{figure}

	Figure~\ref{fig:03} shows the total energy of the $\beta$-Po and the bcc structures of Se as a function of volume. 
       At larger volumes the $\beta$-Po structure is more stable than the bcc structure. 
       At smaller volumes the total energies of both the structures are quite close, but the bcc structure is more stable than the $\beta$-Po structure at volumes smaller than $\sim$ 12~\AA$^3$. 

     In order to estimate the transition pressure from $\beta$-Po to bcc we calculate the Gibbs free energy (or enthalpy) as a function of pressure.
     Then, to evaluate the pressure as a function of volume we fitted the calculated total energies by the Murnaghan's equation of state (EOS)~\cite{40}: 
\[
E(V)=\frac{B_0 V}{B_0'}\left[\frac{1}{B_0'-1}\left(\frac{V_0}{V}\right)^{B_0'}+1\right]+{\rm const},
\]
	where $B_0$ and $B_0'$ is the isothermal bulk modulus at zero pressure and it's derivative, respectively. 
        The pressure is determined from 
\[
P=\frac{B_0}{B_0'}\left[\left(\frac{V}{V_0}\right)^{-B_0'}-1\right].
\]

	The Gibbs free energy is defined by $G(P)\equiv E_{\rm tot}(P)+PV(P)$ and the transition pressure between the two phases is obtained from the relation $G_{\beta}(P)=G_{\rm b}(P)$, where $G_{\beta}$  and $G_{\rm b}$ are the Gibbs free energies of the $\beta$-Po and the bcc structures, respectively. 

	The transition pressure $P_{{\rm c}}$ from $\beta$-Po to bcc has been estimated as 120~GPa by our present calculation.
	This value is higher than other calculated transition pressures, 90~GPa~\cite{38} and 110~GPa,\cite{39} but still lower than the experimental value of 150~GPa.\cite{37}
        The origin of this discrepancy between theory and experiment may be ascribed to the LDA itself and/or numerical accuracy of the total energy. 
        With respect to the latter point we note that Fig.~\ref{fig:03} shows that the volume-energy curves for the two structures are almost parallel near the phase transition. 
        Therefore a small change in the total energy for one of the phases is expected to cause a large change for the value of $P_{\rm c}$; if the total energy of one of the two phases is shifted by 1~mRyd,  the value of $P_c$ changes by 20~GPa. 

\begin{figure}[h]
\begin{center}
\leavevmode
\epsfxsize=80mm
\epsfbox{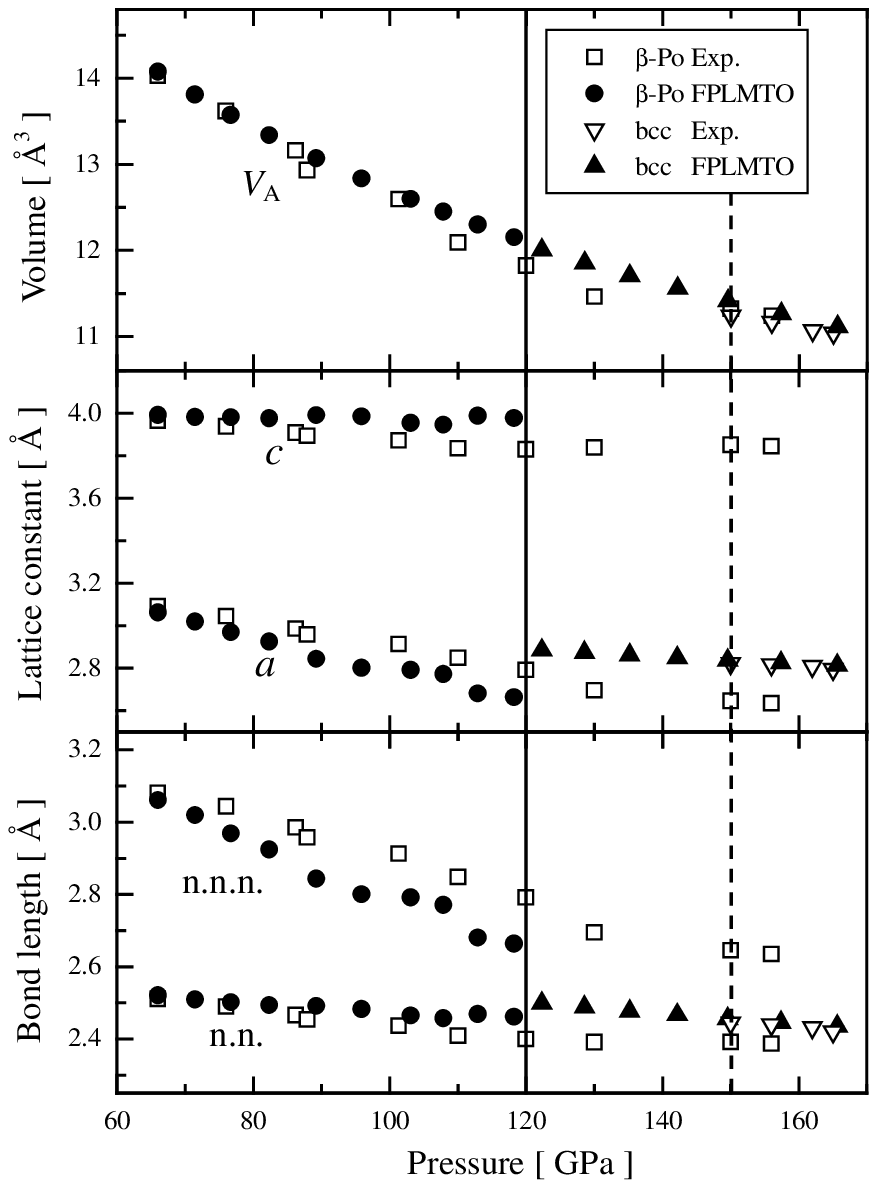}
\caption{The pressure dependence of atomic volume, lattice constants and bond length of selenium. The filed circles and filled squares indicate the results of the present calculations and the open squares and open inverse triangles represent the experimental values.\cite{37} The solid and the dashed vertical lines indicate the boundary of the phase transition from $\beta$-Po to bcc, determined theoretically by us and experimentally by Akahama {\it et.al.}, respectively.}
\label{fig:04}
\end{center}
\end{figure}

	Figure~\ref{fig:04} shows the pressure dependencies of the atomic volume $V_{\rm A}$, the lattice constants $a$ and $c$, and the bond lengths $r_1$ and $r_2$ which are defined as the nearest neighber (n.n.) and next n.n.(n.n.n) atomic distances, respectively. Note that in the bcc phase $c=\frac{\sqrt{6}}{4}a$, and $r_1=r_2=\frac{\sqrt{3}}{2}a$.
        The filled circles and filled squares indicate the results of the present calculations and the open squares and open inverse triangles represent the experimental values.\cite{37}
        The solid and the dashed vertical lines indicate the boundary of the phase transitions determined theoretically by us and experimentally by Akahama {\it et.al.}\cite{37}, respectively.

	As seen from Fig.~\ref{fig:04} the volume variation as a function 
of pressure below 120~GPa ($\beta$-Po) and above 150~GPa (bcc) shows good agreement with the observations.\cite{37} 
	The obtained pressure dependence of $a$, $c$ and the bond lengths 
of the $\beta$-Po phase agree well with the experimental results.\cite{37} 
        Furthermore, the volume reduction at the transition from $\beta$-Po to bcc is estimated to be 0.06~\AA$^3$, which is in good agreement with the experimental volume (about 0.08~\AA$^3$).\cite{37}

\section{LATTICE DYNAMICS, ELECTRON-PHONON INTERACTION AND SUPERCONDUCTIVITY}
\subsection{Calculational procedure}
	Actual calculational procedures are as follows. 
        We find the dynamical matrix as a function of wave vector for a set of irreducible $\bbox{q}$ points on a (8, 8, 8) reciprocal lattice grid [ 29 points in the IBZ ] for the bcc structure and (6, 6, 6) reciprocal lattice grid [ 32 points in the IBZ ] for the $\beta$-Po structure. 
        The ($I$,$J$,$K$) reciprocal lattice grid is defined in a usual manner: $\bbox{q}_{ijk}=(i/I)\bbox{G}_1+(j/J)\bbox{G}_2+(k/K)\bbox{G}_3$, where $\bbox{G}_1, \bbox{G}_2, \bbox{G}_3$ are the primitive translations in the reciprocal space. 

	The self-consistent calculations are performed for every wave vector 
with use of the following basis set and criteria. 
        We use 3$\kappa$--$spd$--LMTO basis set (27 orbitals) with the one-center expansions performed inside the MT spheres up to $\ell_{max}=6$. 
	In the interstitial region the basis functions are expanded in plane waves up to the cutoff corresponding to 134 (110), 176 (170), and 320 (320) plane waves per $s$, $p$, and $d$ orbitals for bcc ($\beta$-Po) structure, respectively.
	The induced charge densities and the screened potentials are represented inside the MT spheres by spherical harmonics up to $\ell_{max}=6$ and in the interstitial region by plane waves with the cutoff corresponding to the (16, 16, 16) fast-Fourier-transform grid in the unit cell of direct space. 
	The $\bbox{k}$-space integration needed for constructing the induced charge density and the dynamical matrix is performed over the (16, 16, 16) grid [ 145 points in the IBZ ] for the bcc structure and (12, 12, 12) grid [ 185 points in the IBZ ] for the $\beta$-Po structure, which is twice denser than the grid of the phonon wave vectors $\bbox{q}$. 
	The integration is performed also by the improved tetrahedron method. 
	However, the integration weights for the $\bbox{k}$ points at these grid have been found to take precisely into account the effects arising from the Fermi surface and the energy bands. 
	This is done with help of the band energies generated by the original FPLMTO method at the (32, 32, 32) grid [ 897 points in the IBZ ] for the bcc structure and (24, 24, 24) grid [ 1313 points in the IBZ] for the $\beta$-Po structure. 
	This procedure allows us to obtain better convergence results with respect to the number of $\bbox{k}$ points. 

	For calculation of the electron-phonon coupling the corresponding $\bbox{k}$-space integrations are more sensitive than the dynamical matrices to the number of sampling $\bbox{k}$-points. 
       It has been performed with the help of the (32, 32, 32) grid for bcc and (24, 24, 24) for $\beta$-Po in the IBZ by means of the tetrahedron method.

	The superconducting transition temperature $T_{\rm c}$ is calculated by using Allen-Dynes formula which is derived on the basis of the strong coupling theory of phonon mechanism. 
	Instead of describing the details of the strong coupling theory, here we give only the necessary equations to calculate $T_{\rm c}$. 
        In the following we completely obey the description of the reference.\cite{13} 

	For the electron-phonon spectral distribution functions $\alpha^2F(\omega)$, we employ the expression~\cite{41} in terms of the phonon linewidths $\gamma_{\bbox{q}\nu}$
\begin{equation}
\label{eqn:spec}
\alpha^2F(\omega)=\frac{1}{2\pi N(\varepsilon_F)}\sum_{\bbox{q}\nu}\frac{\gamma_{\bbox{q}\nu}}{\omega_{\bbox{q}\nu}}\delta(\omega-\omega_{\bbox{q}\nu}),
\end{equation}
where $N(\varepsilon_{\rm F})$ is the electronic density of states per atom and per 
spin at the Fermi level. 
       When the energy bands around the Fermi level are linear in the range of phonon energies, the linewidth is given by the Fermi ``golden rule" and is written as follows:
\begin{equation}
\label{linew}
\gamma_{\bbox{q}\nu}=2\pi\omega_{\bbox{q}\nu}\sum_{\bbox{k}jj'}|g_{\bbox{k}+\bbox{q}j',\bbox{k}j}^{\bbox{q}\nu}|^2\delta(\varepsilon_{\bbox{k}j}-\varepsilon_F)\delta(\varepsilon_{\bbox{k}+\bbox{q}j'}-\varepsilon_F).
\end{equation}
where $g_{\bbox{k}+\bbox{q}j',\bbox{k}j}^{\bbox{q}\nu}$ is the electron-phonon matrix element, and conventionally written in the form
\begin{equation}
\label{eqn:epm}
g_{\bbox{k}+\bbox{q}j',\bbox{k}j}^{\bbox{q}\nu}=\langle\bbox{k}+\bbox{q}j'|\delta^{\bbox{q}\nu}V_{{\rm eff}}|\bbox{k}j\rangle,
\end{equation}
where $\bbox{k}j$ denotes the one-electron basis $\Psi_{\bbox{k}j}$ and $\delta^{\bbox{q}\nu}V_{{\rm eff}}$ is the change in the effective potential induced from a particular $\bbox{q}\nu$ phonon mode. 
       Precisely speaking, the electron-phonon matrix element must be corrected for the incompleteness of the basis functions, but we do not discuss it here.
       The expression of $T_{\rm c}$ derived by Allen-Dynes~\cite{43} by modifying the McMillan formula~\cite{42} is given as 
\begin{equation}
\label{eqn:McM}
T_{\rm c} = \frac{\omega_{{\rm log}}}{1.2}{\rm exp}\left(-\frac{1.04(1+\lambda)}{\lambda - \mu^*(1+0.62\lambda)}\right),
\end{equation}
where 
\begin{equation}
\label{eqn:e-pcoup}
\lambda=2\int_0^{\infty}{\rm d}\omega\frac{\alpha^2F(\omega)}{\omega},
\end{equation}
\begin{equation}
\label{eqn:omegln}
\omega_{{\rm log}}={\rm exp}\frac{1}{\lambda}\int_0^{\infty}\frac{{\rm d}\omega}{\omega}\alpha^2F(\omega){\rm log}\omega.
\end{equation}
	Usually $\lambda$ is called the dimensionless electron-phonon coupling 
constant, $\omega_{{\rm log}}$ the logarithmic -averaged phonon frequency and $\mu^*$ the effective screened Coulomb repulsion constant whose value is usually taken to be between 0.1 and 0.15.

	In case of monatomic metals $\lambda$ can be expressed also in the following form: 
\begin{equation}
\label{eqn:lambda}
\lambda = \frac{N(\varepsilon_{\rm F})\langle I^2\rangle}{M\langle\omega^2\rangle}=\frac{\eta}{M\langle\omega^2\rangle}, 
\end{equation}
where $M$ is the mass of the atoms and $\langle\omega^2\rangle$ denotes the average of squared phonon frequencies which is given as
\begin{equation}
\label{aveomega}
\langle\omega^2\rangle=\frac{\displaystyle\int\omega^2\displaystyle\frac{\alpha^2F(\omega)}{\omega}{\rm d}\omega}{\displaystyle\int\displaystyle\frac{\alpha^2F(\omega)}{\omega}{\rm d}\omega}. 
\end{equation}
Further $\langle I^2 \rangle$ represents the Fermi surface average of squared electron-phonon coupling interaction which is defined by
\begin{equation}
\label{eqn:aveepi}
\langle I^2\rangle=\frac{\displaystyle\sum_{\bbox{q}\nu}\sum_{\bbox{k}jj'}|g_{\bbox{k}+\bbox{q}j',\bbox{k}j}^{\bbox{q}\nu}|^2\delta(\varepsilon_{\bbox{k}j}-\varepsilon_F)\delta(\varepsilon_{\bbox{k}+\bbox{q}j'}-\varepsilon_F)}{\displaystyle\sum_{\bbox{q}\nu}\sum_{\bbox{k}jj'}\delta(\varepsilon_{\bbox{k}j}-\varepsilon_F)\delta(\varepsilon_{\bbox{k}+\bbox{q}j'}-\varepsilon_F)}
\end{equation}
and $\eta$ = $N(\varepsilon_{\rm F})\langle I^2\rangle$ is called the Hopfield parameter.

\subsection{Results for bcc Se}

\begin{figure}[h]
\begin{center}
\leavevmode
\epsfxsize=80mm
\epsfbox{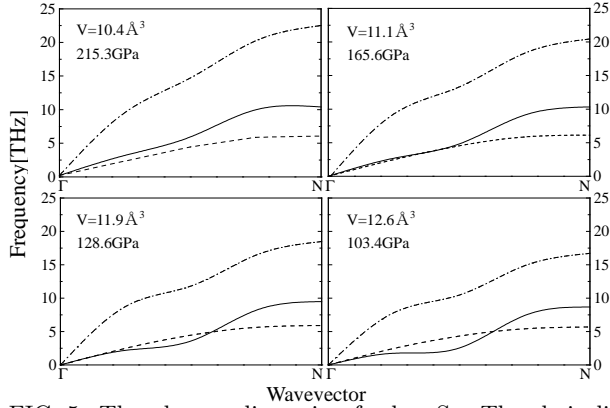}
\caption{The phonon dispersion for bcc Se. The chain line denotes longitudinal mode, and the solid and dashed lines the transverse modes.}
\label{fig:05}
\end{center}
\end{figure}

	We first calculated the phonon dispersion curve along the high symmetry line ($\Gamma$N) for bcc Se at different 4 volumes (pressures), 10.37~\AA$^3$ (214.2~GPa), 11.11~\AA$^3$ (165.6~GPa), 11.85~\AA$^3$ (128.6~GPa) and 12.59~\AA$^3$ (102.59~GPa). 
	The results are shown in Fig.~\ref{fig:05}. 

	As the pressure decreases, the overall tendency of decrease of phonon frequency is seen.
	In particular, the frequency softening is remarkable for one of the transverse modes (shown by the solid curve), and this mode exhibits a notable phonon anomaly, {\it i.e.}, a dip in the middle of the line. 
        The same phonon anomaly is obtained in S.\cite{29}
        This softening of the transverse mode does not cause directly the bcc $\rightarrow$ $\beta$-Po transition with decreasing pressure because both of the $\beta$-Po and bcc phases have one atom per unit cell. However Zakharov and Cohen~\cite{29} have pointed out that it plays an important role in changing the coordination number from eight to six during the bcc $\rightarrow$ $\beta$-Po transition. 

	F. Mauri {\it et al.} have performed {\it ab initio} linear-response 
calculation for lattice dynamics of bcc Te under pressures.\cite{44} 
        They reported the same anomaly for the transverse mode along the 
$\Gamma$N line and found that with decreasing pressure the phonon 
frequencies in the middle of the $\Gamma$N line become imaginary 
in a pressure region where the $\beta$-Po structure is stable. 
	In our calculation complete softening of the transverse mode has 
not been observed even at 100~GPa where the $\beta$-Po structure is stable. 
        Complete softening may occure at even lower pressures. 

\begin{figure}[h]
\begin{center}
\leavevmode
\epsfxsize=80mm
\epsfbox{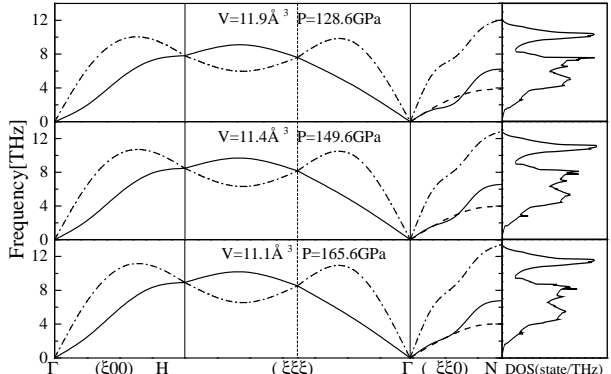}
\caption{The phonon dispersion and phonon density of state (DOS) for bcc Se. The chain line denotes longitudinal modes. The solid and dashed lines denote transverse modes.}
\label{fig:06}
\end{center}
\end{figure}

	Figure~\ref{fig:06} shows the pressure dependence of the phonon dispersion along several symmetry lines and the phonon density of state (DOS) calculated at three volumes (or pressures) 11.85~\AA$^3$ (128.6~GPa), 11.41~\AA$^3$ (149.6~GPa) and 11.11~\AA$^3$ (165.6~GPa). 
        It is noted that except along the $\Gamma$N line all the phonon frequencies soften linearly with decreasing pressure (or increasing volume). 

	By using the Allen-Dynes formula we have estimated the superconducting transition temperature $T_{\rm c}$ of bcc Se at three pressures: 128~GPa, 150~GPa and 166~GPa. 
        In Table~\ref{tab:01} we give the values of calculated $T_{\rm c}$ together with DOS at the Fermi level $N(\varepsilon_{\rm F})$, the Hopfield parameter $\eta$, the logarithmic average frequency $\omega_{{\rm log}}$, the average of squared phonon frequencies $\langle\omega^2\rangle$ and the electron phonon coupling constant $\lambda$. 
        With decreasing pressure the value of $\omega_{{\rm log}}$ decreases while the value of $\lambda$ increases, but the rate of change of $\lambda$ exceeds that of $\omega_{{\rm log}}$.
        As a result the value of $T_{\rm c}$ increases considerably with decreasing pressure.
        Since $\lambda$ can be expressed by 
\[
\lambda = \frac{N(\varepsilon_{\rm F})\langle I^2\rangle}{M\langle\omega^2\rangle}=\frac{\eta}{M\langle\omega^2\rangle}, 
\]
the frequency softening (decrease of $\langle\omega^2\rangle$) is considered to contribute to the increase of $\lambda$ with decreasing pressure. 

\begin{figure}[h]
\begin{center}
\leavevmode
\epsfxsize=80mm
\epsfbox{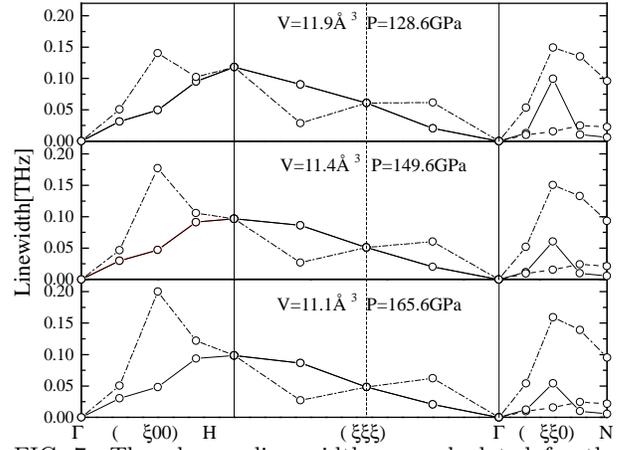}
\caption{The phonon line width $\gamma_{\bbox{q}\nu}$ calculated for three different pressures. The chain line denotes the results for the longitudinal mode, and the solid and dashed ones those for the transverse modes. Note that the transverse modes are degenerate along the $\Gamma$-H-P-$\Gamma$ line.}
\label{fig:07}
\end{center}
\end{figure}

        In order to obtain a more physical insight into the characteristic pressure dependence of $T_{\rm c}$ we consider mode and wave-vector dependencies of the phonon linewidths $\gamma_{\bbox{q}\nu}$ along the symmetry lines. 
        Figure \ref{fig:07} shows that $\gamma_{\bbox{q}\nu}$ is almost independent of pressure except for the longitudinal mode along the $\Gamma$H line and one of the transverse modes along the $\Gamma$N.

       With decreasing pressure, $\gamma_{\bbox{q}\nu}$ of the longitudinal mode along the $\Gamma$H decreases whereas that of the transverse mode along the $\Gamma$N line increases considerably. 
       Generally speaking, a large phonon linewidth increases the dimensionless electron-phonon coupling $\lambda$.
       Therefore, it is expected that the transverse mode along the $\Gamma$N line plays an important role in giving rise to the characteristic pressure dependence of $T_{\rm c}$. 

\begin{figure}[h]
\begin{center}
\leavevmode
\epsfxsize=80mm
\epsfbox{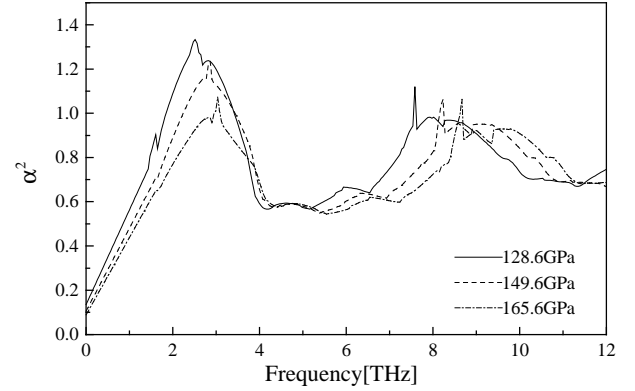}
\caption{The frequency dependence of $\alpha (\omega)^2$ obtained for three pressures.}
\label{fig:08}
\end{center}
\end{figure}

         To clarify the role of the transverse mode along the $\Gamma$N line in more detail we have calculated a quantity $\alpha^2(\omega)$ defined by
\begin{equation}
\label{eqn:alpha}
\alpha^2(\omega)=\frac{\alpha^2F(\omega)}{D(\omega)}, 
\end{equation}
where $\alpha^2F(\omega)$ is the spectral function and $D(\omega)$ denotes the phonon density of states. 
        We consider that by inspecting the frequency dependence of $\alpha^2(\omega)$ we can discern which phonons make dominant contributions to the dimensionless electron-phonon coupling $\lambda$. 
        Fig~\ref{fig:08} shows the calculated $\alpha^2(\omega)$ as a function of frequency for three pressures. 
       The peak around 2~THz originates from transverse phonons along the $\Gamma$N line and the peak around 7$\sim$10~THz from longitudinal phonons along the $\Gamma$H line. 
        As seen from the figure, both the peaks move towards the lower frequency side with decreasing pressure. 
        It should be noted, however, that the magnitude of $\alpha^2(\omega)$ around 2~THz increases remarkably with decreasing pressure whereas the magnitude of $\alpha^2(\omega)$ around 7$\sim$10~THz is less dependent on pressure. 
        Therefore, we can say again that transverse phonons in the middle of the $\Gamma$N line make a dominant contribution to $\lambda$.

       Combining all of the above results we conclude that the origin of remarkable increase of $T_{\rm c}$ of bcc Se with decreasing pressure is mainly attributed to the phonon anomaly (the remarkable frequency softening) in the middle of the $\Gamma$N line.


\subsection{Results for $\beta$-Po Se}

\begin{figure}[h]
\begin{center}
\leavevmode
\epsfxsize=80mm
\epsfbox{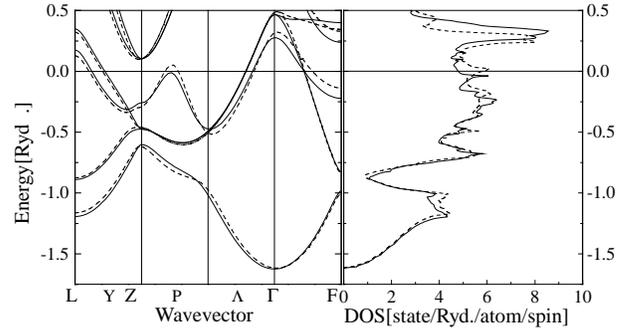}
\caption{The band structure and the DOS of $\beta$-Po Se at atomic volume $V$=12.59~\AA\ (103.1 GPa.). The dashed curves denote the results obtained with use of lattice constants estimated by calculation and the solid ones those with use of lattice constants determined by experiments.\cite{37} The horizontal line denotes the Fermi level.}
\label{fig:09}
\end{center}
\end{figure}

       In this section we calculate the superconducting transition temperature of $\beta$-Po Se. 
       To see the pressure dependence of $T_{\rm c}$ in $\beta$-Po Se we have calculated $T_{\rm c}$ at pressures 103.1~GPa and 118.2~GPa with two sets of lattice constants: one is the lattice constants evaluated by calculation and the other is those determined by experiment and given in Table~\ref{tab:02}. 
       Figure~\ref{fig:09} shows the electronic dispersion curves and the density of states calculated for 103.1~GPa. 

       The calculated $T_{\rm c}$ are given in Table~\ref{tab:03} together with the electronic DOS at the Fermi level $N(\varepsilon_{\rm F})$, the Hopfield parameter $\eta$, the logarithmic average frequency $\omega_{{\rm log}}$, the average of squared phonon frequencies $\langle\omega^2\rangle$ and the electron phonon coupling constants $\lambda$. 
       $T_{\rm c}$ strongly depends on which sets of lattice constants is used. 
       For the lattice constants estimated by calculation' the magnitude of $T_{\rm c}$ is larger and increases considerably with decreasing pressure. 
       For lattice constants determined by measurements, on the other hand, the magnitude of $T_{\rm c}$ is smaller and depends little on pressure. 

       The logarithmic average frequencies $\omega_{{\rm log}}$ are larger for the experimental lattice constants. 
       The electron-phonon coupling $\lambda$, on the other hand, is larger for the theoretical lattice constants and furthermore depends considerably on pressure, which gives higher and more pressure-sensitive transition temperatures $T_{\rm c}$ for the theoretical lattice constants.  

\begin{figure}[h]
\begin{center}
\leavevmode
\epsfxsize=80mm
\epsfbox{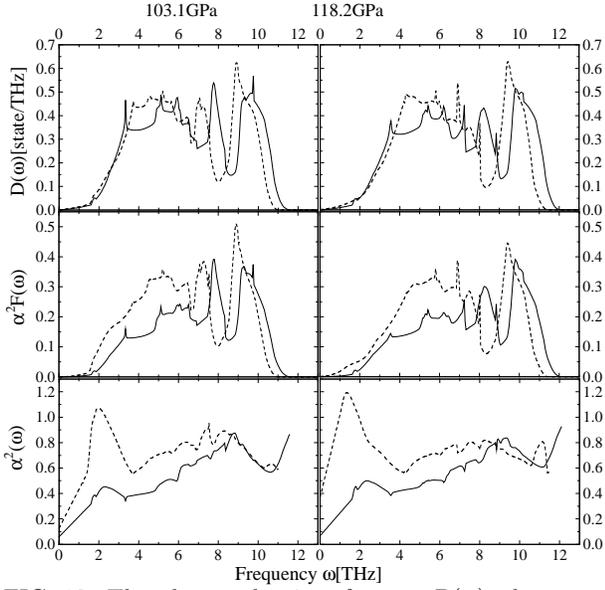}
\caption{The phonon density of states $D(\omega)$, the spectral function $\alpha^2F(\omega)$ and $\alpha^2(\omega)$ defined by Eq.~\ref{eqn:alpha} calculated for pressures 103.1~GPa (left hand side) and 118.2~GPa (right hand side) with use of the theoretical and experimental lattice constants.  The dashed curves represent the results for the theoretical lattice constants and the solid curves for the experimental lattice constants.}
\label{fig:10}
\end{center}
\end{figure}

	In order to clarify the origin of the different magnitude and the different pressure dependence of $T_{\rm c}$ for different sets of lattice constants we have calculated the phonon density of states $D(\omega)$, the spectral function $\alpha^2F(\omega)$ and $\alpha^2(\omega)$ defined by Eq.~\ref{eqn:alpha}.
	The results are shown in Fig.~\ref{fig:10}. 
	The magnitude of $\alpha^2(\omega)$ for the theoretical lattice constants is larger than that for the experimental lattice constants throughout almost the whole frequency range.
	In particular the magnitude of a peak in $\alpha^2(\omega)$ around 2~THz obtained for the calculational lattice constants is remarkably enhanced compared to that obtained for the experimental ones, thus the magnitude of $\lambda$ and $T_{\rm c}$ take large values.
	For the experimental lattice constants, $\alpha^2(\omega)$ depends little on pressure and hence the magnitude of $\lambda$ and $T_{\rm c}$ are also less dependent on pressure. 
	
\begin{figure}[h]
\begin{center}
\leavevmode
\epsfxsize=80mm
\epsfbox{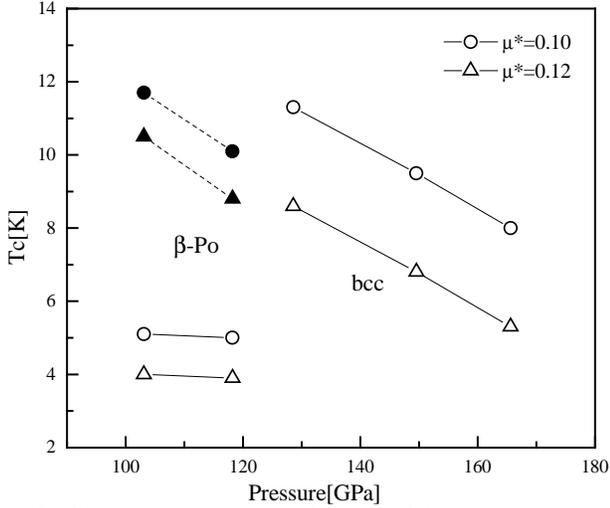}
\caption{The superconducting transition temperature $T_{\rm c}$ of Se as a function of pressure.  The closed circles and closed triangles denote the computed values of $T_{\rm c}$ with $\mu^*$=0.12 and $\mu^*$=0.10, respectively.  For the $\beta$-Po structure the open circles and open triangles represent the results for the lattice constants estimated by calculation and the closed circles and closed triangles those for the lattice constants determined by experiments.}
\label{fig:11}
\end{center}
\end{figure}

	Fig.~\ref{fig:11} shows the values of $T_{\rm c}$ calculated for $\beta$-Po and bcc Se as a function of pressure. 
	If we adopt the experimental lattice constants for $\beta$-Po Se, the superconducting transition temperature $T_{\rm c}$ is almost pressure independent in the $\beta$-Po phase and there is a large jump in $T_{\rm c}$ at the transition from $\beta$-Po to bcc.  

\section{CONCLUSION}
	We have performed FPLMTO calculation for the $\beta$-Po and the bcc 
structures of selenium, and succeeded in reproducing the phase transition 
from $\beta$-Po to bcc as observed by experiments. 
        The obtained pressure dependencies of lattice parameters agree 
fairly well with the experimental results. 
        The estimated transition pressure is 120~GPa. 
        It improves the previous theoretical calculations, but is still lower 
than the experimental value 150~GPa.

	We have investigated the superconductivity of $\beta$-Po and 
bcc Se by calculating the lattice dynamics and electron-phonon interaction 
with use of LR-FPLMTO method. 
	For bcc Se we have found that the frequency softening is remarkable 
for one of the transverse modes in the middle of the $\Gamma$N-line, 
and this mode exhibits a notable phonon anomaly.
	The calculated superconducting transition temperature $T_{\rm c}$ 
increases considerably with decreasing pressure, 
which is mainly attributed to the phonon anomaly in the middle of 
the $\Gamma$N line. 
        In the $\beta$-Po structure, on the other hand, $T_{\rm c}$ is 
smaller compared to that of the bcc phase and is almost independent of pressure.
        Finally we predicte a discontinuous jump in $T_{\rm c}$ 
at the transition from $\beta$-Po to bcc.

\acknowledgments
 	We thank greatly Dr. S.Yu. Savrazov of Max Planck Institute for providing us with the FPLMTO and LR-FPLMTO program.
	We are grateful to Prof. Y. Akahama for showing us their experimental results prior to publications.
	One of the author (M.O.) thanks Prof. M. Shirai for illuminating and valuable discussion about linear-response theory and also Dr. K. Yamaguchi for fruitful discussion about FPLMTO method.
	This work is partly supported by Grant-in-Aid for COE Reaearch (10CE2004), Grant-in-Aid for Scientific Research (C)(09640433) of the Ministry of Education, Science, Sports and Culture.
	     M.O. is supported by the JSPS Research Fellowships 
for Young Scientists. \\

%
%

%
%
\begin{table}[h]
\caption{The electronic DOS at the Fermi level $N(\varepsilon_{\rm F})$, the Hopfield parameter $\eta$, the logarithmic average frequencies $\omega_{{\rm log}}$, the average of squared phonon frequencies $\langle\omega^2\rangle$, the electron phonon coupling constant $\lambda$ and the superconducting transition temperatures $T_{{\rm c}}$ calculated as a function of pressure for bcc Se.  The two values for $T_{{\rm c}}$ correspond to two different values of $\mu^*$ (0.10 and 0.12). The units of $N(\varepsilon_{\rm F})$ and $\eta$ are state/Ryd./atom/spin and Ryd./a$^2$, respectively.} 
\label{tab:01}
\begin{center}
\begin{tabular}{c@{\extracolsep{\fill}}cccccc}
$P$ (GPa)  & $N(\varepsilon_{\rm F})$ & $\eta$ & $\omega_{{\rm log}}$ (K) & $\langle\omega^2\rangle$ (K$^2$) & $\lambda$ & $T_{\rm c}$ (K) \\ \hline
128.6 & 5.45 & 0.20 & 224.73 & 291.95$^2$ & 0.83 & 11.29, 9.90 \\
149.6 & 5.24 & 0.21 & 248.03 & 316.41$^2$ & 0.73 &  9.53, 8.11 \\ 
165.6 & 5.10 & 0.21 & 264.62 & 335.11$^2$ & 0.66 &  8.03, 6.64
\end{tabular}
\end{center}
\end{table}

\begin{table}[h]
\caption{Two sets of lattice constants $c/a$ and $a$ of $\beta$-Po Se determined by calculation and experiments.}
\label{tab:02}
\begin{center}
\begin{tabular}{ccccc}
 [GPa]   & 103.1 calc. & 103.1 Exp. & 118.2 calc. & 118.2 Exp. \\ \hline
      c/a & 0.71  & 0.75  & 0.67  & 0.74  \\
 a (\AA)  & 7.471 & 7.314 & 7.504 & 7.255
\end{tabular}
\end{center}
\end{table}

\begin{table}[h]
\caption{The electronic DOS at the Fermi level $N(\varepsilon_{\rm F})$, the Hopfield parameter $\eta$, the logarithmic average frequencies $\omega_{{\rm log}}$, the average of squared phonon frequencies $\langle\omega^2\rangle$, the electron phonon coupling constant $\lambda$ and the superconducting transition temperatures $T_{{\rm c}}$ of $\beta$-Po Se calculated for 103.1~GPa and 118.2~GPa.  The upper two lines show the results obtained with use of the lattice constants estimated by calculation and the lower two lines those with use of the lattice constants determined by experiments.  The two values for $T_{{\rm c}}$ correspond to two different values of $\mu^*$ (0.10 and 0.12).  The units of $N(\varepsilon_{\rm F})$ and $\eta$ are state/Ryd./atom/spin and Ryd./a$^2$, respectively. 
}
\label{tab:03}
\begin{center}
\begin{tabular}{c@{\extracolsep{\fill}}cccccc}
$P$ (GPa.)  & $N(\varepsilon_F)$ & $\eta$ & $\omega_{{\rm log}}$ (K) & $\langle\omega^2\rangle$ (K$^2$) & $\lambda$ & $T_{{\rm c}}$ (K) \\ \hline
103.1 & 5.23 & 0.19 & 192.98 & 264.72$^2$ & 0.92 & 11.74, 10.49 \\
118.2 & 5.50 & 0.19 & 204.84 & 279.76$^2$ & 0.82 & 10.10,  8.84 \\ \hline
103.1 & 4.89 & 0.16 & 250.11 & 311.71$^2$ & 0.58 &  5.14,  4.04 \\
118.2 & 4.85 & 0.17 & 255.88 & 324.58$^2$ & 0.57 &  5.01,  3.91
\end{tabular}
\end{center}
\end{table}

\end{document}